\documentclass[conference]{IEEEtran}
\usepackage[dvips]{graphicx}
\usepackage{indentfirst}
\usepackage{amsmath}
\usepackage{amssymb}
\usepackage{threeparttable}
\usepackage{multirow}
\usepackage{subfigure}
\usepackage{xcolor}
\usepackage{cite}
\usepackage{enumerate}
\usepackage{algorithm}
\usepackage{algorithmic}
\usepackage{setspace}

\newtheorem{Theorem}{Theorem}

\newtheorem{Lemma}{Lemma}

\ifCLASSOPTIONcompsoc
\else
\fi

\ifCLASSINFOpdf
\else
\fi

\allowdisplaybreaks

\begin{document}

\markboth{Submitted to IEEE GLOBECOM 2018}{ Submitted to IEEE GLOBECOM 2018}

\title{Resource Allocation for Low-Latency Vehicular Communications with Packet Retransmission}

\author{Chongtao Guo$^\dag$, Le Liang$^{\ddag}$, and Geoffrey Ye Li$^{\ddag}$ \\
$^\dag$ College of Information Engineering, Shenzhen University, Shenzhen 518060, Guangdong, China \\
$\ddag$ School of Electrical and Computer Engineering, Georgia Institute of Technology, Atlanta, GA 30332, USA\\
    Email: ctguo@szu.edu.cn; lliang@gatech.edu; liye@ece.gatech.edu
}

\maketitle

\begin{abstract}

Vehicular communications have  stringent latency requirements on safety-critical information transmission.
However,  lack of instantaneous channel state information due to high mobility poses a great challenge to meet these requirements and the situation gets more complicated when packet retransmission is considered.
Based on only the obtainable large-scale fading channel information, this paper performs spectrum and power allocation to maximize the ergodic capacity of vehicular-to-infrastructure (V2I) links while guaranteeing the latency requirements of vehicular-to-vehicular (V2V) links.
First, for each possible spectrum reusing pair of a V2I link and a V2V link, we obtain the closed-form expression of the packets'  average  sojourn time (the queueing time plus the service time) for the V2V link.
Then, an optimal power allocation is derived for each possible spectrum reusing pair.
Afterwards, we optimize the spectrum reusing pattern by addressing a polynomial time solvable bipartite matching problem.
Numerical results show that the proposed queueing analysis is accurate in terms of  the  average  packet sojourn time.
Moreover, the developed resource allocation always guarantees the V2V links' requirements on latency.

\end{abstract}


\IEEEpeerreviewmaketitle

\section{ Introduction }\label{Section_Introduction}

Recent development of autonomous driving and on-wheel infortainment services have accelerated the interest in  vehicular communications.
Vehicular communications include vehicle-to-vehicle (V2V), vehicle-to-infrastructure (V2I), and vehicle-to-pedestrian (V2P) communications, which are collectively referred to as  vehicle-to-everything (V2X) communications \cite{2017-TVT-VehCommunPhysicalLayerPers, 2017-arXiv-VehCommunNetLayer}.
While the V2I links often require high-capacity data delivery, the V2V links have stringent demand on reliability and latency since the V2V links often exchange  safety-critical  information,  such as common awareness messages (CAM) and decentralized notification messages (DENM) \cite{2017-MVT-CoopITSinEurope}.
The IEEE 802.11p based vehicular communications have been widely studied in recent years.
However, its carrier-sense based multiple access scheme faces great challenges in guaranteeing strict   quality-of-service (QoS) requirements of V2X communications, especially when the traffic load grows heavy \cite{2017-MVT-LTEVsidelink5GV2X}.
As an alternative, the cellular assisted vehicular communications with QoS-aware resource allocation have a sufficient potential to guarantee the diverse QoS requirements of  different types of links, where the V2V and V2P communications are performed based on the cellular-assisted device-to-device (D2D) technique \cite{2014-MCOM-D2DCommunInCellular}.

The D2D technique enables proximate users to communicate directly with each other, which leads to the proximity, hop, and reuse gains \cite{2014-MCOM-D2DCommunInCellular}.
To largely exploit these gains, most of the existing studies have preferred the reuse mode to the dedicated mode \cite{2014-TCOM-JointModeSelRA}, where the reuse mode allows the D2D users to share the cellular users' spectrum and  the dedicated mode assigns exclusive spectrum to the D2D users \cite{2013-TCOM-D2DCommunUnderlayingCellular}.
Traditional studies on D2D communications cannot be directly applied to vehicular communications due to the perfect channel state information (CSI) assumed available at the base stations (BS) or the D2D transmitters.
This assumption does not hold any more since the channel varies fast owning to the high mobility of vehicles and it is quite difficult, if not impossible, to estimate and feed the instantaneous CSI back to the transmitters.
To this end, the D2D-enabled vehicular communications should carefully address  the challenge caused by the channel uncertainty.
The work in \cite{2017-WCL-SpectrumPowerAlloWithDelayCSI} have considered the case of delayed feedback of the CSI of the V2V links and proposed a spectrum and power allocation scheme to maximize the sum capacity of the V2I links with guarantee on the V2V links' signal-to-interference-plus-noise ratio (SINR) outage probability.
Resource management in \cite{2017-TCOM-ResourceAllocationForD2DV2X} have maximized the sum and minimum ergodic capacity of the V2I links while guaranteeing the SINR outage probabilities of the V2V links, based on only the large-scale channel information.
Considering requirements on the reliability and transmission latency of the V2V links,  resource allocation schemes have been developed in  \cite{2016-TWC-ClusterBasedRAManageForD2DV2X}  and \cite{2016-TVT-RAManageForD2DBasedV2V} for the cases of permitting and not permitting spectrum sharing among the V2V links, respectively.


The overall end-to-end latency of a communication link contains four parts \cite{2017-arXiv-LowLatencyMMWaveCommun, 1992-Book-DataNetworks},  i.e., processing latency, propagation latency, transmission latency, and queueing latency.
When the packet retransmission is considered in case of an unsuccessful transmission, the queueing latency and the equivalent transmission latency are the dominant parts while the other two ones can be comparatively negligible \cite{2017-arXiv-LowLatencyMMWaveCommun}.
However, most of the existing works have only considered the transmission latency for the V2V links while neglecting the dominant queueing latency.
In this paper, we take into consideration the different QoS requirements of the V2I and V2V links, i.e., the V2I links pursue high-capacity data transmission and the V2V links require stringent latency.
The latency is described by the average  packet sojourn time, i.e., the time of waiting for transmission in the queue plus the time of transmission and retransmission.
In addition, we will assume only the large-scale slowly varying channel fading is available for resource allocation since the fast varying component is quite difficult to obtain.
The proposed spectrum and power allocation is to maximize the sum ergodic capacity of the V2I links while guaranteeing the latency of the V2V links.
First, the latency expression is derived by analyzing the queueing system for each possible V2I-V2V spectrum reusing pair under a fixed power allocation.
Then, the optimal power allocation for each possible V2I-V2V pair is obtained based on the queueing analysis results.
Finally, the spectrum reusing pattern is optimized by addressing a  bipartite matching problem.


\section{System Model}\label{Section_SystemModel}

Consider a single cell with the BS located at its center, where  $M$ vehicles, called CUEs, are performing V2I communications and $K$ ($K\le M$) pairs of vehicles, called DUEs, require high-reliability low-latency V2V communications.
Assume that  $M$  mutually orthogonal spectrum bands have been assigned to the $M$ CUEs and each DUE needs to reuse the spectrum of one CUE.
Time is divided into slots with length $T$, on the order of hundreds of microseconds, and $L$ consecutive slots form a block, on the order of hundreds of milliseconds.
Since the large-scale fading of  channels is typically determined by vehicle locations, which are less likely to change too much during one block, it is assumed   invariant  during one block but may change from one block to another.
However, due to the high mobility of vehicles, we assume independent and identically distributed small-scale channel components for different slots, which remain constant during one slot.
The BS has the knowledge of the large-scale fading information of all channels, which can be estimated or fed back from the DUEs once a block with low signaling overhead.
As for the small-scale fading components, we assume the BS only knows their distributions instead of their realizations, since it is impractical to feed full CSI back to the BS in every slot.
In the rest of this paper, we investigate the BS-side centric power and spectrum allocation issue based on the large-scale fading information for a particular block.
Let $P_m^c$ and $P_k^d$  denote the transmit powers of the $m$th CUE and the $k$th DUE, respectively.
The notation $x_{m,k}\in\{0,1\}$ is the spectrum reusing pattern indicator regarding the $m$th CUE and the $k$th DUE, where $x_{m,k}=1$ indicates that the $k$th DUE reuses the spectrum of the $m$th CUE and $x_{m,k}=0$ otherwise.
We assume that the spectrum of each CUE can be reused by at most one DUE and each DUE can share the spectrum of at most one CUE, i.e., $\sum_{k=1}^K x_{m,k}\le 1$ and $\sum_{m=1}^M x_{m,k}\le 1$.

The channel power gain from the $m$th CUE to the BS, denoted by $h_{m}^c$, is formulated as $h_m^c = \alpha_m^c g_m^c$, where $\alpha_m^c$ and $g_m^c$ represent the slowly varying large-scale fading and the fast varying small-scale fading, respectively.
The large-scale fading $\alpha_m^c$ is assumed to follow $\alpha_m^c = G \beta_m^c (d_m^c)^{-\varphi}$, where $G$ is the pathloss constant, $d_m^c$ is the distance between the $m$th CUE and the BS, $\varphi$ is the power decay exponent, and $\beta_m^c$ is a log-normal shadow fading random variable with a standard deviation $\xi$.
The channel of the $k$th DUE, the crosstalk channel from the $k$th DUE to the BS, and the crosstalk channel from the $m$th CUE to the $k$th DUE are similarly defined as $h_k^d = \alpha_k^d g_k^d$, $h_{k,B}=\alpha_{k,B}g_{k,B}$, and $h_{m,k}=\alpha_{m,k}g_{m,k}$, respectively.
In this paper, we consider the Rayleigh fast fading for all channels, i.e.,  $g_m^c$, $g_k^d$, $g_{k,B}$, and $g_{m,k}$ are independent and exponentially distributed with unit mean for different slots.

The transmitter of each DUE has an infinite-size buffer to store the arriving and backlogged packets.
Consider that the modulation has been determined and thus the data transmission rates over channels are constants.
The packets have equal length corresponding to the number of  bits that can be transmitted during one slot, i.e., one and only one packet can be transmitted during one slot.
The packets arriving in the queue at the transmitter of the $k$th DUE follow  the Poisson arrival process with intensity $\lambda_k$ and  they are served (transmitted)  according to the rule of first-come-first-served (FCFS).
We assume $\lambda_k T < 1$ always holds, which is a necessary condition for guaranteeing a  stable queue \cite{1992-Book-DataNetworks,2006-Book-QueueNetworks}.
The automatic repeat-request (ARQ) protocol is considered and thus the success or failure (outage) of the packet detection at the receiver is fed back to the transmitter through an instantaneous error-free control channel \cite{2011-TCOM-CodARQinFading, 2014-TMC-ThrOptWirelessNetworks, 2018-TWC-StableThrAndDelayAna}.
If the packet at the head-of-line (HOL) of the queue has not been successfully transmitted during the current slot, it will be retransmitted in the next slot and this process continues until it is successfully received at the receiver.
We can see that there is no packet loss here and the reliability can be surely guaranteed for the DUEs.
Let  $\mu_k^d$ denote the  average  packet sojourn time of the $k$th DUE  that accounts for the queueing time plus the service time.

With an assumption that the CUEs always have data to be delivered, the  SINR of the $k$th DUE, denoted by $\gamma_k^d$, is
\begin{equation}
\gamma_k^d = \frac{P_k^d \alpha_k^d g_k^d}{\sigma^2 + \sum_{m=1}^M x_{m,k} P_m^c \alpha_{m,k} g_{m,k}},
\end{equation}
where $\sigma^2$ is the  power of the additive white Gaussion noise.
In an arbitrary slot, the packet of the $k$th DUE is unsuccessfully transmitted with probability ${\rm Pr}\{\gamma_k^d < \gamma_0 \}$, where $\gamma_0$ is the minimum required SINR  for a successful packet detection.
Even if the $k$th DUE shares the spectrum of the $m$th CUE, it does not always pose interference to the $m$th CUE.
This is because the channel will not be used by the $k$th DUE in some slots  if the queue is empty at the beginning of these slots.
Thus, with $\rho_k^d$ denoting the channel busy probability of the  $k$th DUE, the ergodic capacity of the $m$th CUE, denoted by $R_m^c$, can be expressed as in (\ref{eqn_ErgCapacityRmc}).
\begin{table*}
\begin{equation}\label{eqn_ErgCapacityRmc}
\begin{split}
R_m^c  = & \sum_{k=1}^K x_{m,k} \left\{
           (1-\rho_k^d) \mathbb{E} \left[  \log_2 \left( 1 + \frac{P_m^c \alpha_m^c g_m^c}{\sigma^2} \right) \right]
          +     \rho_k^d  \mathbb{E}  \left[ \log_2 \left( 1 + \frac{P_m^c \alpha_m^c g_m^c}{\sigma^2 +  P_k^d \alpha_{k,B} g_{k,B} } \right)   \right]
          \right\}
\end{split}
\end{equation}
\rule{\textwidth}{0.2mm}
\end{table*}

Recognizing different QoS requirements for different types of links, we maximize the sum ergodic capacity of the CUEs while guaranteeing the  latency  for each DUE.
In addition, we provide a minimum guaranteed QoS for each CUE by setting a requirement on the minimum achievable ergodic capacity.
The ergodic capacity of the CUEs  can be computed through the long-term average over the fast fading.
In particular, faster variation implies more slots contained within a given block and thus the system time average throughput approaches the computed ergodic capacity quicker \cite{2017-TCOM-ResourceAllocationForD2DV2X}.
To this end, the spectrum and power allocation problem is formulated as
\begin{subequations}
\label{OriginalProblem}
\begin{eqnarray}
 \underset{   \{P_m^c\},  \{P_k^d\}, \{x_{m,k}\}      }{\max}  &  \sum_{m=1}^M R_m^c  \label{OriginalProblem_Object}  \\
   {\rm s.t.}  \ \ \ \ \ \ \ \          &    \mu_k^d \le \mu_0,   \forall k                  \label{OriginalProblem_Con1}  \\
              &    R_m^c \ge R_0,  \forall m                       \label{OriginalProblem_Con2}  \\
              &    0 \le P_k^d \le P_{\max}^d,  \forall k               \label{OriginalProblem_Con3} \\
              &    0 \le P_m^c \le P_{\max}^c,   \forall m                \label{OriginalProblem_Con4} \\
              &    \sum_{m=1}^M x_{m,k} \le 1,  \forall k                \label{OriginalProblem_Con5} \\
              &    \sum_{k=1}^K x_{m,k} \le 1,  \forall m              \label{OriginalProblem_Con6} \\
              &    x_{m,k}\in \{0,1\},  \forall m,k,                 \label{OriginalProblem_Con7}
\end{eqnarray}
\end{subequations}
where  $\mu_0$ and $P_{\max}^d$ are  the maximum allowed average  packet sojourn time  and the maximum allowed transmit power for the DUEs, respectively, and $R_0$ and $P_{\max}^c$ are the minimum required ergodic capacity and the maximum allowed transmit power for the CUEs, respectively.

The main challenge in solving  problem (\ref{OriginalProblem}) lies in the coupling spectrum and power allocation variables.
In addition,   constraint  (\ref{OriginalProblem_Con1}) and the objective function are involved with the DUEs' queueing performance, i.e., the average  packet sojourn time and the channel busy probabilities, which have not been explicitly expressed.
To this end, we separate the power and spectrum allocation by observing that interference exists only between each CUE-DUE spectrum reusing pair.
For each possible CUE-DUE pair, we design the optimal power allocation to maximize the CUE's ergodic capacity while guaranteeing the DUE's latency.
After that, we check the feasibility of each spectrum reusing pair against the CUE's minimum capacity requirement, rule out infeasible pairs, and construct a bipartite matching problem to find the optimal spectrum reusing pattern.

To assure that constraint (\ref{OriginalProblem_Con1}) is satisfied, we first need to find the expression of the average  packet sojourn time of the DUEs.
In the next section, we will study the queueing performance for each CUE-DUE spectrum reusing pair under a fixed power allocation.

\section{Queueing Analysis} \label{Section_QueueAnalysis}

In this section, we analyze the queueing system of the DUE by considering a spectrum reusing pair of the $m$th CUE and the $k$th DUE.
As the time has been divided into slots, a new packet arriving in an empty queue   should wait for the end of the idle slot rather than be  served immediately.
In other words, the server (i.e., the wireless channel) goes on vacation for some random interval of time.
Then, the behavior of the packets at the transmitter of the  DUE can be characterized by the $M/G/1$ queueing with vacations \cite{1992-Book-DataNetworks}, where $M$, $G$, and $1$ indicate that packets arrive in the queue  according to the Poisson process, the service time of packets are generally distributed, and there is only one server, respectively.
It is worth pointing out that the queueing system should be kept stable, i.e., the queueing length will not go to infinity as the time approaches infinity.
Otherwise, the average  packet sojourn time   will go to infinity and the latency requirement of the V2V link cannot be guaranteed.
The following lemma, proved in \cite{1992-Book-DataNetworks}, presents a sufficient and necessary condition of  stable $M/G/1$ queueing systems and the average  sojourn time of customers for a stable $M/G/1$ queueing system.

\begin{Lemma}\label{Lemma_MG1Basic}
For an $M/G/1$ queueing system with server vacations with $\lambda$ being the Possion  arrival intensity of customers and $Y$ being the random service time of a customer, the server busy probability, denoted by $\rho$, is given by $\rho= \lambda \mathbb{E}\{Y\}$
and the queueing system is stable if and only if $\rho<1$.
In addition,  the average  sojourn time of customers for a stable $M/G/1$ queueing system with vacations, denoted by $\mu$,  is
\begin{equation}\label{eqn_MG1vacaLatency}
\mu =  \mathbb{E}\{Y\} + \frac{\lambda \mathbb{E}\{Y^2\}}{2(1-\lambda \mathbb{E}\{Y\})} +  \frac{\mathbb{E}\{V^2\}}{2\mathbb{E}\{V\}} ,
\end{equation}
where $V$ is the random vacation interval of the server.
\end{Lemma}

The stability condition in Lemma \ref{Lemma_MG1Basic} shows that  on  average  the number  of  packets  generated in  a unit  of  time must be  less than the number  of  packets  that  can be processed \cite{1992-Book-DataNetworks}.
This condition should  be always satisfied  as the results derived in this paper only apply to stable systems.
Noticing that  the server takes a vacation for one slot when there is no packet in the queue at the beginning of a slot, we have the first and second moments of the vacation interval $V$ given by $\mathbb{E}\{V\}=T$ and $\mathbb{E}\{V^2\}=T^2$, respectively.

The probability that a packet of the $k$th DUE is unsuccessfully transmitted during an arbitrary slot, denoted by $q_k$, is characterized by the SINR outage probability given by

\vspace{-0.5em}
\begin{small}
\begin{equation}\label{eqn_OutPro_q}
\begin{split}
q_k & = {\rm Pr}\left\{ \frac{P_k^d \alpha_k^d g_k^d}{\sigma^2 +  P_m^c \alpha_{m,k} g_{m,k}}  < \gamma_0 \right\}
      = 1 - \frac{P_k^d \alpha_k^d \exp\left(-\frac{\gamma_0 \sigma^2}{P_k^d \alpha_k^d}\right)}{P_k^d \alpha_k^d + \gamma_0 P_m^c \alpha_{m,k}},
\end{split}
\end{equation}
\end{small}

\vspace{-0.5em}
\noindent where the second equality follows from \cite[Lemma 1]{2017-TCOM-ResourceAllocationForD2DV2X}.
The random service time of a packet of the $k$th DUE is denoted by $Y_k$, which has the probability mass function given by
\begin{equation}\label{eqn_YkPMF_InfN}
{\rm Pr}\{Y_k=jT\} = q_k^{j-1}(1-q_k),   j = 1,2,\cdots.
\end{equation}
The first and second moments of $Y_k$ will be
\begin{equation}
\begin{split}
\mathbb{E}\{Y_k\}  = \sum_{j=1}^{\infty} jT q_k^{j-1}(1-q_k) = \frac{T}{1-q_k},
\end{split}
\end{equation}
and
\begin{equation}
\begin{split}
    \mathbb{E}\{Y_k^2\}  = \sum_{j=1}^{\infty} (jT)^2 q_k^{j-1}(1-q_k)
= \frac{T^2(1+q_k)}{(1-q_k)^2},
\end{split}
\end{equation}
respectively.

According to Lemma \ref{Lemma_MG1Basic}, the channel busy probability of the $k$th DUE is
\begin{equation}\label{eqn_ServerBusyPro_Case3}
\rho_{k}^d = \lambda_k \mathbb{E}\{Y_k\} =\frac{\lambda_k  T}{1-q_k},
\end{equation}
which implies that $q_k< 1-\lambda_k T$ is the sufficient and necessary condition for guaranteeing a stable queue.
The average  packet sojourn time of the $k$th DUE for a stable queueing system can then be given by
\begin{equation}\label{eqn_SojournTime_Case3}
\mu_{k}^d = \frac{T}{2} +  \frac{T}{1-q_k} + \frac{\lambda_k T^2 (1+q_k)  }{2(1-q_k)(1-q_k-\lambda_k T )}.
\end{equation}
It can be found that the latency $\mu_{k}^d$ monotonically increases with $q_k$ in the  region $(0,1-\lambda_k T)$.
Then, the infimum of the achievable average  packet sojourn time, denoted by $\mu_{k,\min}$, shows up at  $q_k=0$ as given as
\begin{equation}
\mu_{k,\min} = \frac{T(3-2\lambda_k T)}{2(1-\lambda_k T)}.
\end{equation}
On the other hand, $\mu_{k}^d$ tends to be infinite as $q_k$ approaches  $1-\lambda_k T$.
Therefore, if $\mu_0 > \mu_{k,\min}$, there must exist $\overline{q}_k \in (0, 1-\lambda_k T)$ such that the latency constraint (\ref{OriginalProblem_Con1}) is equivalent to $q_k   \le \overline{q}_k$,
where $\overline{q}_k$ is the solution to the equation $\mu_k^d = \mu_0$.
Otherwise, the latency requirement of the $k$th DUE cannot be satisfied and problem (\ref{OriginalProblem}) is infeasible.

To derive the closed-form expression for $\overline{q}_k$ in case of feasibility, we transform the equation $\mu_k^d = \mu_0$ to a quadratic equation given by
\begin{equation}
\begin{split}
(2\mu_0-T)\overline{q}_k^2 - [2\lambda_k T^2 -2(\lambda_k \mu_0 + 2)T + 4 \mu_0]\overline{q}_k \\
+ [2\lambda_kT^2 - 2(\lambda_k \mu_0 + 1 )T - (T-2\mu_0)] =0.
\end{split}
\end{equation}
One root of the above equation is $1$, which is out of our interest, and the other root is
\begin{equation}\label{eqn_Overlineqk}
\overline{q}_k = \frac{2\lambda_k T^2-(2\lambda_k \mu_0+3)T + 2\mu_0}{2\mu_0-T},
\end{equation}
which lies in the region $(0,1)$ as declared in what follows.
First, from $\mu_0 > \mu_{k,\min}=\frac{T(3-2\lambda_k T)}{2(1-\lambda_k T)}>T$, we can see $2\lambda_k T^2-(2\lambda_k \mu_0+3)T + 2\mu_0 >0$ and thus $\overline{q}_k>0$.
Second, since the numerator minus the denominator leads to $2T(\lambda_k T - \lambda_k \mu_0 -1)<0$, we have $\overline{q}_k<1$.

\section{Proposed Resource Allocation}\label{Section_ProposedAlgorithm}

In this section, we first perform the optimal power allocation for each possible spectrum reusing pair and then determine the optimal spectrum reusing pattern.

\subsection{Power Allocation for Each CUE-DUE Pair}

The power allocation problem for the spectrum reusing pair of the $m$th CUE and the $k$th DUE can be formulated as
\begin{subequations}
\label{PowerAlloProblem}
\begin{eqnarray}
 \underset{  P_k^d , P_m^c }{\max}  &   R_{m}^c( P_k^d , P_m^c )   \label{PowerAlloProblem_Object}  \\
   {\rm s.t.} \   &    q_k( P_k^d , P_m^c ) \le \overline{q}_k                \label{PowerAlloProblem_Con1}  \\
              &    0 \le P_k^d \le P_{\max}^d              \label{PowerAlloProblem_Con2} \\
              &    0 \le P_m^c \le P_{\max}^c,                 \label{PowerAlloProblem_Con3}
\end{eqnarray}
\end{subequations}
where $\overline{q}_k$ is given in (\ref{eqn_Overlineqk}),  $q_k(P_k^d,P_m^c)$ is defined as
\begin{equation}
q_k( P_k^d , P_m^c ) = 1 - \frac{P_k^d \alpha_k^d \exp\left(-\frac{\gamma_0 \sigma^2}{P_k^d \alpha_k^d}\right)}{P_k^d \alpha_k^d + \gamma_0 P_m^c \alpha_{m,k}},
\end{equation}
and $R_{m}^c( P_k^d , P_m^c )$ is expressed by

\vspace{-0.5em}
\begin{small}
\begin{equation}\label{eqn_ObjPairMK}
\begin{split}
    &  R_{m}^c( P_k^d , P_m^c )  \\
=  & \left[1- \frac{\lambda_k T}{1-q_k(P_k^d,P_m^c)}\right] \cdot \mathbb{E} \left\{  \log_2 \left( 1 + \frac{P_m^c \alpha_m^c g_m^c}{\sigma^2} \right) \right\}  \\
  + & \frac{\lambda_k T}{1-q_k(P_k^d,P_m^c)}  \cdot  \mathbb{E}  \left\{ \log_2 \left( 1 + \frac{P_m^c \alpha_m^c g_m^c}{\sigma^2 +  P_k^d \alpha_{k,B} g_{k,B} } \right)   \right\}.
\end{split}
\end{equation}
\end{small}

\noindent Notice that  constraint (\ref{PowerAlloProblem_Con1}) ensures the latency requirement of the $k$th DUE, i.e., $\mu_k^d\le \mu_0$, according to the previous section.
The constraint of the CUE's ergodic capacity requirement is dropped here and will be reconsidered in the spectrum allocation.
The following theorem, proved in Appendix \ref{Appendix_Theorem_OptSolutionValue}, characterizes the optimal solution to  problem (\ref{PowerAlloProblem}).
\begin{Theorem}\label{Theorem_OptSolutionValue}
The optimal power allocation to  problem (\ref{PowerAlloProblem}) is given by
\begin{equation}\label{eqn_OptPower}
{P_k^d}^* = \min\{P_{\max}^d, f^{-1}(P_{\max}^c)\},   {P_m^c}^* = \min\{P_{\max}^c, f(P_{\max}^d)\},
\end{equation}
where $f(\cdot)$ is defined as
\begin{equation}
 f(P_k^d) \triangleq \frac{\alpha_k^d P_k^d}{\gamma_0 \alpha_{m,k}} \left( \frac{1}{1-\overline{q}_k}\exp\left(-\frac{\gamma_0 \sigma^2}{P_k^d \alpha_k^d}\right) - 1 \right),
\end{equation}
and the optimal objective value is

\vspace{-0.5em}
\begin{small}
\begin{equation}\label{eqn_OptObjValue}
\begin{split}
 & R_{m}^c({P_k^d}^*,{P_m^c}^*) = \left[1- \frac{\lambda_k T}{1-q_k( {P_k^d}^* , {P_m^c}^* )}\right] \cdot \frac{{\rm e}^{\frac{1}{a}} E_1(\frac{1}{a})}{\ln 2}   \\
 & + \frac{\lambda_k T}{1-q_k( {P_k^d}^* , {P_m^c}^* )} \cdot \frac{a\left[{\rm e}^{\frac{1}{a}} E_1(\frac{1}{a}) - {\rm e}^{\frac{1}{b}} E_1(\frac{1}{b})  \right]}{(a-b)\ln 2},
\end{split}
\end{equation}
\end{small}

\noindent where $a=\frac{{P_m^c}^* \alpha_m^c}{\sigma^2}$, $b=\frac{{P_k^d}^* \alpha_{k,B}}{\sigma^2}$, and $E_1(x)=\int_{x}^{\infty} \frac{{\rm e}^{-\zeta}}{\zeta} {\rm d} \zeta$.

\end{Theorem}

Theorem \ref{Theorem_OptSolutionValue} yields the optimal power allocation for a single CUE-DUE pair, where the value $f^{-1}(P_{\max}^c)$ can be found by bisection searching since the function $f(\cdot)$ is monotonically increasing in the region of interest.
The rest is to conduct spectrum reusing pattern optimization based on the optimal power allocation for each possible CUE-DUE reusing pair.

\subsection{Spectrum Allocation}

Let $R_{m,k}^*$ denote the optimal ergodic capacity of the $m$th CUE if the $k$th DUE reuses the spectrum of the $m$th CUE.
For the spectrum reusing pair of the $m$th CUE and the $k$th DUE,  the optimal power allocation $({P_k^d}^*,{P_m^c}^*)$ can be obtained from Theorem \ref{Theorem_OptSolutionValue}, which leads to the maximum capacity of the $m$th CUE, denoted by $R_{m}^c({P_k^d}^*,{P_m^c}^*)$,  while guaranteeing the latency requirement of the $k$th DUE.
Although $R_{m}^c({P_k^d}^*,{P_m^c}^*)$ is the maximum achievable capacity, it might be still less than the CUE's minimum required capacity  $R_0$ since the constraint on the CUE's capacity has been removed in the proposed power allocation.
In this case,  the reusing pair is infeasible.
To take the minimum required capacity as constrained in (\ref{OriginalProblem_Con2})  into consideration, we define
\begin{equation}
R_{m,k}^* =
\begin{cases}
R_{m}^c({P_k^d}^*,{P_m^c}^*), & {\rm if} \ R_{m}^c({P_k^d}^*,{P_m^c}^*) \ge  R_0 \\
-\infty,                              & {\rm otherwise},
\end{cases}
\end{equation}
in the spectrum reusing pattern optimization.
The spectrum reusing pattern optimization issue can then be formulated as
\begin{subequations}
\label{ProblemMatching}
\begin{eqnarray}
 \underset{   \{ x_{m,k} \}  }{\max}  \   &    \sum_{m=1}^M \sum_{k=1}^K x_{m,k} R_{m,k}^*                \label{ProblemMatching_Object}  \\
  {\rm s.t.}    \ \                    &    \sum_{m=1}^M x_{m,k} \le 1, \forall k           \label{ProblemMatching_Con1}  \\
                                       &    \sum_{k=1}^K x_{m,k} \le 1, \forall m           \label{ProblemMatching_Con2}  \\
                                       &    x_{m,k} \in \{0,1\},  \forall k, m,                           \label{ProblemMatching_Con3}
\end{eqnarray}
\end{subequations}
which is a weighted bipartite matching problem and can be efficiently  addressed by the Hungarian method \cite{2001-Book-GraphTheory}.

In summary, the proposed resource allocation is globally optimal since the reusing pattern is optimally chosen from all possible reusing pairs based on their optimal power allocations.
We now discuss the complexity of the proposed method.
First, the complexity of the power allocation for all $KM$ possible CUE-DUE pairs  is $\mathcal{O}(KM \log(1/\epsilon))$, where $\epsilon$ is the error tolerance of the bisection searching.
Second, the Hungarian method solves the spectrum reusing pattern optimization problem in $\mathcal{O}(M^3)$ time.
As a result, the overall complexity of the proposed algorithm is $\mathcal{O}(KM\log(1/\epsilon) + M^3)$.

\begin{figure}[!t]
\centering
  \includegraphics[width=7.5cm]{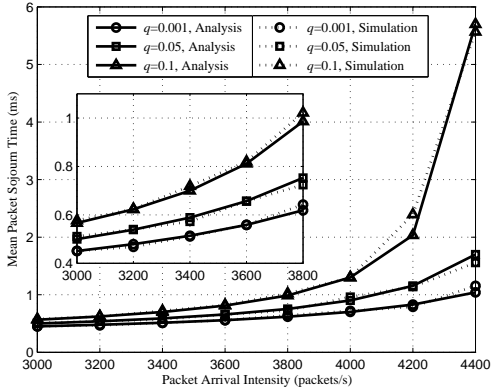}
  \caption{\label{lambdaVSlatency_analysisVerify} DUE's average  packet sojourn time  vs. DUE's packet arrival intensity. }
\end{figure}

\section{Numerical Results}\label{Section_SimuResults}


Considering the $k$th DUE sharing the spectrum of one CUE, we verify the accuracy of the proposed queueing analysis by simulating its queueing system  for a block of $L=20,000$ time slots with slot length $T=0.2$ ms under varying  SINR outage probability  $q_k$.
As shown in Fig. \ref{lambdaVSlatency_analysisVerify},  the analytical average  packet sojourn time coincides with the simulation results for different SINR outage probabilities and packet arrival intensities.
It can also be observed that the latency grows with the packet arrival intensity  since the increasing number of packets entering the queue will lengthen the queue and the packets need to wait more time for transmission.
In addition, a smaller $q_k$ leads to shorter latency since the packets will less likely to need retransmission.

\begin{figure}
\centering
  \includegraphics[width=7.2cm]{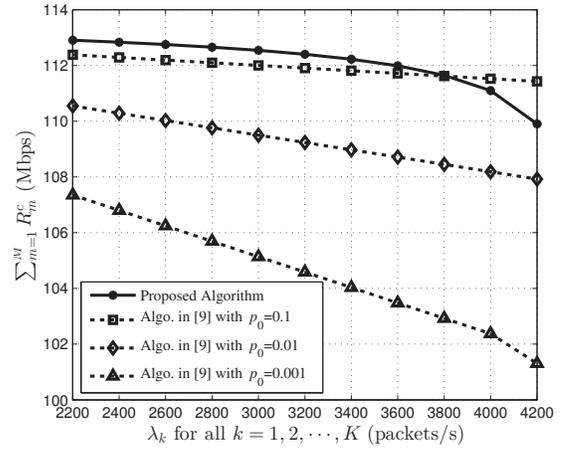}
  \caption{\label{EffeConfirm_lambdaVScapacity} CUEs' sum ergodic capacity  vs.  DUEs' packet arrival intensity.
  }
\end{figure}
\begin{figure}
\centering
  \includegraphics[width=7.2cm]{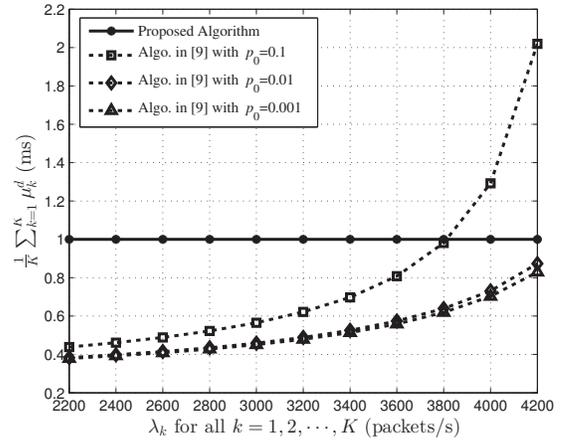}
  \caption{\label{EffeConfirm_lambdaVSlatency} DUEs' average  packet sojourn time  vs. DUEs' packet arrival intensity.
  }
\end{figure}

We now perform the proposed resource allocation and the algorithm in \cite{2017-TCOM-ResourceAllocationForD2DV2X} in a randomly generated network with $M=20$ CUEs and $K=20$ DUEs, where vehicles move at an average speed 60 km/h, $T=0.2$ ms, $P_{\max}^c=P_{\max}^d=23$ dBm, $R_0=0.5$ Mbps,  $\gamma_0 = 5$ dB, $\mu_0=1$ ms,  and other parameters are the same as \cite{2017-TCOM-ResourceAllocationForD2DV2X}.
The SINR outage probability requirement in \cite{2017-TCOM-ResourceAllocationForD2DV2X} is denoted by $p_0$.
Fig. \ref{EffeConfirm_lambdaVScapacity} and Fig. \ref{EffeConfirm_lambdaVSlatency} present the sum ergodic capacity of the CUEs and the latency of the DUEs, respectively.
Since the latency of the DUEs is overly guaranteed by the strategy in  \cite{2017-TCOM-ResourceAllocationForD2DV2X} when $p_0=0.01$ and $p_0=0.001$, it has much lower sum capacity for the CUEs than the proposed algorithm.
When $p_0=0.1$, the scheme in  \cite{2017-TCOM-ResourceAllocationForD2DV2X} results in higher capacity for the CUEs when the packet arrival intensity is greater than 3800 packets/s, however, at the cost of failing to guarantee the latency for the DUEs.
In summary,  there is a tradeoff between the QoS of the CUEs and the DUEs, and the capacity of the CUEs should be improved only on the premise of guarantee on timely delivery of the safety-critical information for the DUEs.

\section{Conclusion}\label{Section_Conclusion}
The developed resource allocation has maximized the sum ergodic capacity of the V2I links while guaranteeing the latency requirements of the V2V links in vehicular networks.
First, we have analyzed the latency of the V2V links  and  provided the optimal power allocation for each V2I-V2V spectrum reusing pair.
Then, the spectrum reusing pattern has been optimized by addressing a bipartite matching problem using the Hungarian algorithm.
Numerical results have verified the accuracy of the proposed queueing analysis and confirmed the effectiveness of the developed resource allocation.

\section{Acknowledgment}\label{Section_Acknowledgment}

The work of C. Guo was supported in part by the National Natural Science Foundation of China under Grant 61601307, in part by Foundation of Shenzhen under Grant JCYJ20160422102022017, and in part by China Scholarship Council (CSC) under Grant 201708440022.
The work of L. Liang and G. Y. Li was supported in part by a research gift from Intel Corporation and in part by the National Science Foundation under Grants 1443894 and 1731017.



\appendices

\section{Proof of Theorem \ref{Theorem_OptSolutionValue}}\label{Appendix_Theorem_OptSolutionValue}

Given a feasible power allocation $(P_k^d , P_m^c)$, we now show that $R_{m}^c(P_k^d , P_m^c)$ will increase if we  enlarge $P_k^d$ and $P_m^c$ with the same factor $\theta$ ($\theta>1$) while still keep $\theta P_k^d \le P_{\max}^d$ and $\theta P_m^c \le P_{\max}^c$.
To see this, we express $R_{m}^c(P_k^d,  P_m^c)$ as
\begin{equation}
\begin{split}
R_{m}^c(P_k^d,  P_m^c) =   R_{m,1}^{c}(P_k^d,  P_m^c) + R_{m,2}^{c}(P_k^d,  P_m^c),
\end{split}
\end{equation}
where $R_{m,1}^{c}(P_k^d,  P_m^c)$ and $R_{m,2}^{c}(P_k^d,  P_m^c)$ are defined as

\vspace{-0.5em}
\begin{small}
\begin{equation}
\begin{split}
R_{m,1}^{c}(P_k^d,  P_m^c) & =   \mathbb{E} \left\{\log_2\left( 1 + \frac{P_m^c \alpha_m^c g_m^c}{\sigma^2 + P_k^d \alpha_{k,B} g_{k,B}} \right)\right\} 
\end{split}
\end{equation}
\end{small}
and
\begin{small}
\begin{equation}
\begin{split}
  & R_{m,2}^{c}(P_k^d,  P_m^c)= \\
& \left[1- \frac{\lambda_k T}{1-q_k(P_k^d, P_m^c)} \right]
\mathbb{E} \left\{\log_2\left(
\frac{1+\frac{P_m^c \alpha_m^c g_m^c}{\sigma^2}}
{1+\frac{P_m^c \alpha_m^c g_m^c}{\sigma^2 + P_k^d \alpha_{k,B} g_{k,B}}}
\right)\right\}
\end{split}
\end{equation}
\end{small}

\vspace{-0.5em}
\noindent respectively.
With the observation that $q_k(\theta P_k^d, \theta P_m^c) < q_k(P_k^d, P_m^c)$, it can be found that the power allocation $(\theta P_k^d , \theta P_m^c)$ is feasible, $R_{m,1}^{c}(\theta P_k^d, \theta P_m^c) > R_{m,1}^{c}(P_k^d, P_m^c)$, and $R_{m,2}^{c}(\theta P_k^d, \theta P_m^c) > R_{m,2}^{c}(P_k^d, P_m^c)$, where the detailed proof is omitted here due to space limitation.
Then, we have $R_{m}^c(\theta P_k^d , \theta P_m^c) > R_{m}^c(P_k^d , P_m^c)$, which indicates that one can always improve the objective value by enlarging $P_k^d$ and $P_m^c$ with the same factor until one of them reaches its maximum allowed value.
Further, since problem (\ref{PowerAlloProblem}) has the same constraints with problem (6) in \cite{2017-TCOM-ResourceAllocationForD2DV2X}, we can  derive the optimal solution given  in (\ref{eqn_OptPower}) by following  \cite[Appendix B]{2017-TCOM-ResourceAllocationForD2DV2X}.

The ergodic capacity of  the $m$th CUE in  case of suffering  no interference is

\vspace{-0.5em}
\begin{small}
\begin{equation}\label{eqn_ErgCapacityNoInterference}
\begin{split}
  &  \mathbb{E}\left\{ \log_2 \left( 1 + \frac{P_m^c \alpha_m^c g_m^c}{\sigma^2 } \right)  \right\} \\
= & \int_0^{\infty} \log_2 \left( 1 + \frac{P_m^c \alpha_m^c g_m^c}{\sigma^2 } \right) \cdot {\rm e}^{-g_m^c} {\rm d} g_m^c = \frac{ {\rm e}^{\frac{1}{a}}E_1(\frac{1}{a})}{\ln 2}, \\
\end{split}
\end{equation}
\end{small}

\vspace{-0.5em}
\noindent where $a=\frac{P_m^c \alpha_m^c}{\sigma^2}$, $E_1(x)=\int_x^{\infty} \frac{{\rm e}^{-\zeta}}{\zeta} {\rm d}\zeta$, and the last equation follows from \cite[eq.(3.352.4)]{2007-Book-TableOfIntegrals}.
Then, following the similar steps as that in \cite[Appendix C]{2017-TCOM-ResourceAllocationForD2DV2X},  we can derive the optimal objective value given in (\ref{eqn_OptObjValue}).


\bibliographystyle{IEEEtran}
\bibliography{IEEEabrv,Reference_V2XwithRetransmission}

\end{document}